\documentclass[Physsubmission, Phys]{SciPost}

\usepackage[T1]{fontenc}
\usepackage[english]{babel}
\usepackage{amssymb,amsmath}
\usepackage{graphicx}
\usepackage[font=scriptsize]{subcaption}
\usepackage[font=small]{caption}
\usepackage{float}
\usepackage{pgf}
\usepackage{tikz}
\usepackage{physics}
\usetikzlibrary{arrows.meta}
\usepackage{mathtools}
\usepackage{sidecap}
\usepackage{xfrac}
\usepackage{tikz-feynman}
\usepackage{authblk}
\usepackage[symbol]{footmisc}

\hypersetup{
    colorlinks,
    linkcolor={red!50!black},
    citecolor={blue!50!black},
    urlcolor={blue!80!black}
}

\usepackage[bitstream-charter]{mathdesign}
\DeclareSymbolFont{usualmathcal}{OMS}{cmsy}{m}{n}
\DeclareSymbolFontAlphabet{\mathcal}{usualmathcal}

\begin{document}

\begin{center}{\Large \textbf{
Local Unitarity\\
}}\end{center}

\begin{center}
Zeno Capatti\textsuperscript{1}*
\end{center}

\begin{center}
{\bf 1} ETH Zürich, Rämistrasse 101, 8092 Zürich, Switzerland
\\
* zcapatti@phys.ethz.ch
\end{center}

\begin{center}
\today
\end{center}


\definecolor{palegray}{gray}{0.95}
\begin{center}
\colorbox{palegray}{
  \begin{tabular}{rr}
  \begin{minipage}{0.1\textwidth}
    \includegraphics[width=35mm]{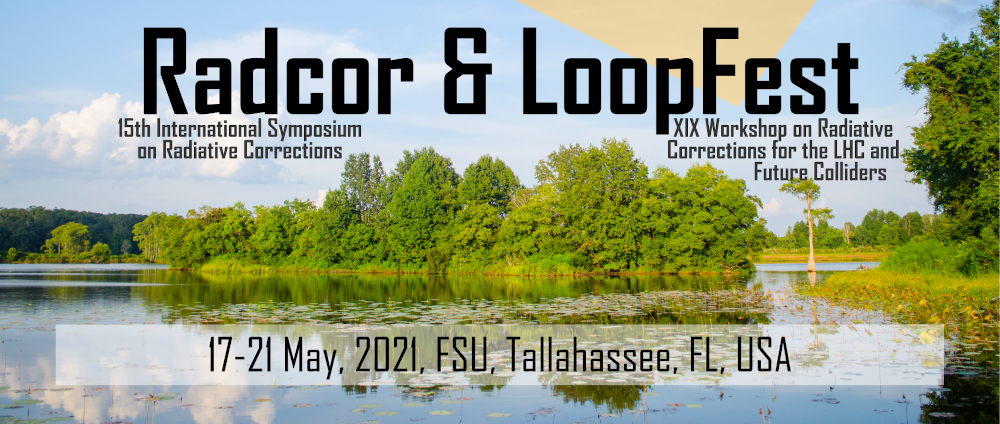}
  \end{minipage}
  &
  \begin{minipage}{0.85\textwidth}
    \begin{center}
    {\it 15th International Symposium on Radiative Corrections: \\Applications of Quantum Field Theory to Phenomenology,}\\
    {\it FSU, Tallahasse, FL, USA, 17-21 May 2021} \\
    \doi{10.21468/SciPostPhysProc.?}\\
    \end{center}
  \end{minipage}
\end{tabular}
}
\end{center}

\section*{Abstract}
{\bf
Within the Local Unitarity formalism, any physical cross-section is re-written in such a way that cancellations of infrared singularities between real and virtual contributions are realised locally. Consequently, phase-space and loop integrals are regulated locally without the need of any counter-terms or dimensional regularisation aside from those strictly needed to achieve ultraviolet regularisation. This Local Unitarity representation is especially suitable to direct Monte Carlo integration. As such, it offers a clear and direct path to automate the computation of fixed-order differential observables in Quantum Field Theories.
}

\vspace{10pt}
\noindent\rule{\textwidth}{1pt}
\tableofcontents\thispagestyle{fancy}
\noindent\rule{\textwidth}{1pt}
\vspace{10pt}

\section{Motivation: forward-scattering diagrams}
\label{sec:intro}
The simplest available practical application of the Local Unitarity formalism is the $\gamma^\star\rightarrow d \bar d$ process at Next-To-Leading-Order in QCD. The S-matrix has a perturbative representation as a power series in a coupling $g$. Each coefficient of the series can be represented diagrammatically. To order $\mathcal{O}(g_\mathrm{s}^2)$ in the strong coupling, the S-matrix can be written diagrammatically as

\begin{minipage}{13.52cm}
\begin{figure}[H]
\centering
\includegraphics[width=13.5cm]{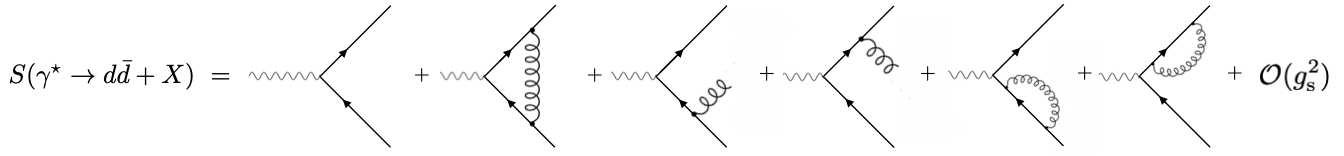}
\end{figure}
\end{minipage}%
\begin{minipage}{1cm}
\begin{center}
\vspace{0.2cm}
  (1)  
\end{center}
\end{minipage}
\vspace{0.25cm}

Each of the diagrammatic contributions appearing in eq.~$(1)$ corresponds to a function of the external momenta (each corresponding to the half-edges of the graph) that can be written explicitly using Feynman rules. As an example, one may consider the diagram

\begin{minipage}{13.52cm}
\begin{figure}[H]
\centering
\includegraphics[width=7cm]{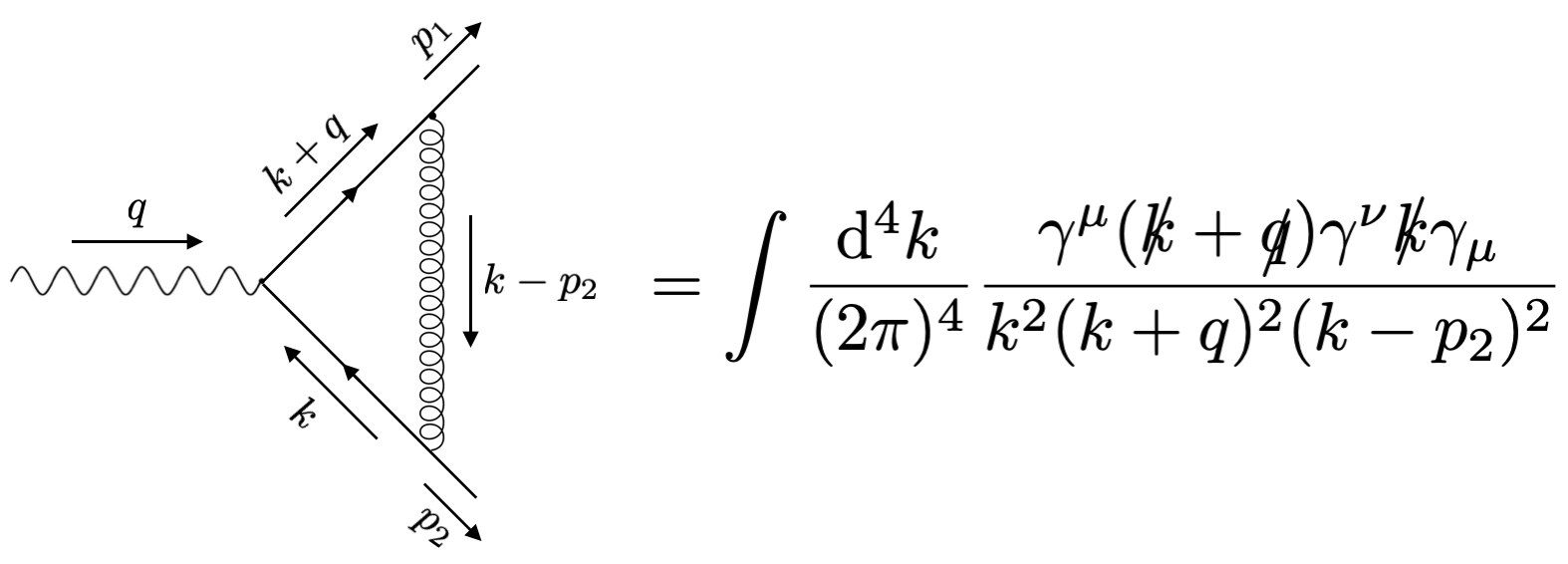}
\end{figure}
\end{minipage}%
\begin{minipage}{1cm}
\begin{center}
\vspace{0.2cm}
  (2)  
\end{center}
\end{minipage}

\vspace{0.3cm}
\noindent expressed as a parametric four-dimensional integral, or loop integral, whose value depends on the external momenta $q$, $p_1$ and $p_2$. From now on, we will assume that the system is in the rest frame of the massive decaying particles, $p_1^2=p_2^2=0$ and $q=(\sqrt{q^2},0,0,0)$. The integral in eq.~$(2)$ is ill-defined, as it features two types of singularities: a) the logarithmic singularity in the limit $k\rightarrow \infty$, also called a UV singularity and b) the logarithmic singularities in the limits $k\rightarrow -x p_2$ and $k\rightarrow -q +x p_1$, with $x\in[0,1]$, known as infrared (IR) singularities.

While a systematic approach to the regularisation of UV singularities is guaranteed by the BPHZ procedure~\cite{Bogolioubov,Hepp,Zimmermann}, no equivalently generic treatment is available for IR singularities. Conversely, strict constraints on the structure and origin of IR singularities allow to conclude that physical quantities, that are observable in collider experiments, should be free of such singularities. This is known as the KLN theorem~\cite{Kinoshita:1962ur,Lee:1964is}. The mechanism that ensures this principle is commonly known as that of "real-virtual cancellations". This name has its origin in the type of contributions that add up to define physical observables.

More precisely, a physical cross-section is obtained by squaring the S-matrix, and integrating over the phase-space of external particles. This yields a power series in the QCD coupling $\alpha_\mathrm{s}$. The diagrammatic depiction of these contributions is:

\begin{minipage}{13.52cm}
\begin{figure}[H]
\centering
\includegraphics[width=13cm]{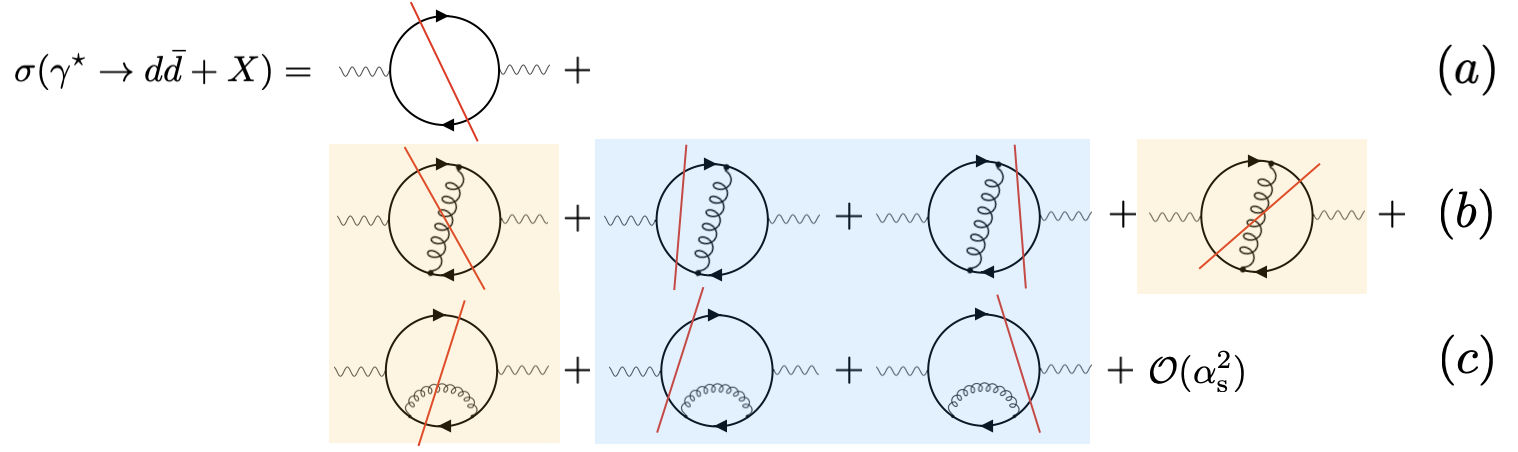}
\end{figure}
\end{minipage}%
\begin{minipage}{1cm}
\begin{center}
\vspace{0.2cm}
  (3)  
\end{center}
\end{minipage}
\vspace{0.25cm}

Diagrams like those exhibited in rows $a$, $b$ or $c$ are called interference diagrams. Each interference diagram identifies an underlying forward-scattering diagram, which is the graph obtained from the interference diagram by ignoring the red cut. It follows that there is a one-to-one correspondence between rows of eq.~$(3)$ and forward-scattering diagrams. 

Line $a$ constitutes the Leading-Order contribution, corresponding to the zeroth order term of the perturbative series, while $b$ and $c$ constitute the Next to Leading-Order contributions, corresponding to first order term of the perturbative series. Conventionally, the $\mathrm{NLO}$ diagrams which (do not) feature a closed loop after deletion of the cut edges is known as a (real) virtual contribution and are highlighted in blue (yellow). At $\mathrm{N}^k\mathrm{LO}$, interference diagrams are classified by the number of loops that are left after deletion of the edges crossed by the cut. 

Each particle traversed by the red cut (also known as Cutkosky cut) is set to be on-shell, and its momentum is thus integrated over its mass-shell. Focusing on one of the virtual contributions of eq.~(3), this rule explicitly translates into a double integral, also known as a phase-space integral,

\begin{minipage}{13.52cm}
\begin{figure}[H]
\centering
\includegraphics[width=13cm]{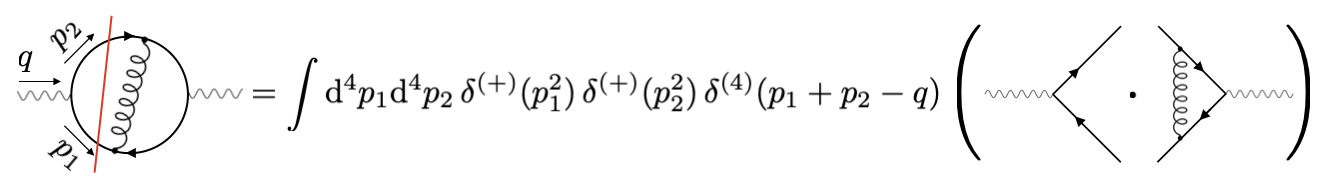}
\end{figure}
\end{minipage}%
\begin{minipage}{1cm}
\begin{center}
\vspace{0.2cm}
  (4)  
\end{center}
\end{minipage}
\vspace{0.25cm}

\noindent The product of diagrams being integrated over can itself be written in terms of integrals analogously to what was done in eq.~(2).

The key statement of the KLN theorem is encoded in the following observation: while each interference diagram has IR singularities, $\sigma$ does not. An even stronger statement can be put forward: each row ($a$, $b$ or $c$) of eq.~(3) is \textit{separately} IR-finite, both inclusively and differentially. The easiest way to see that this holds, at the \textit{inclusive} and \textit{integrated} level, utilizes the optical theorem. For example, row $(b)$ can be rewritten as

\begin{minipage}{13.52cm}
\begin{figure}[H]
\centering
\includegraphics[width=13cm]{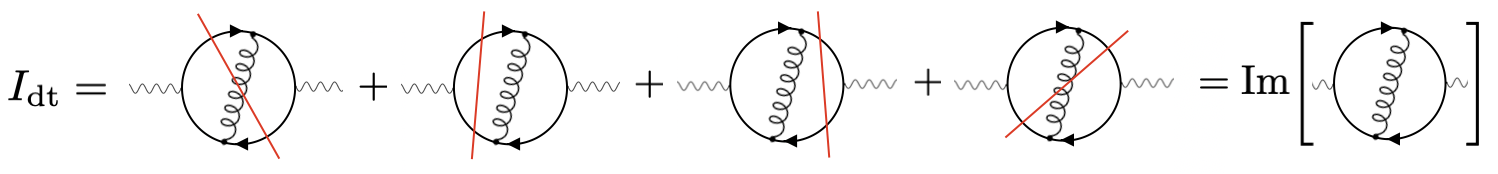}
\end{figure}
\end{minipage}%
\begin{minipage}{1cm}
\begin{center}
\vspace{0.2cm}
  (5)  
\end{center}
\end{minipage}
\vspace{0.25cm}

\noindent and the right hand side, the imaginary part of the double-triangle forward-scattering diagram is manifestly IR-finite for $q^2>0$. This principle turns out to be very general:

\vspace{0.5cm}
\textbf{The sum over all Cutkosky cuts of a forward-scattering diagram (with arbitrary number of initial states) is free of final state singularities}.
\vspace{0.5cm}

\section{Local Unitarity}

Given that such a constraining and natural cancellation mechanism exists for infrared singularities, it is worth asking if it is possible to realize it locally. That is: given a physical, potentially differential cross-section $\sigma$, is it possible to express the coefficients of the perturbative expansion of an observable as
\begin{equation}\tag{6}
    \sigma=\sum_{L=1}^\infty \alpha_\mathrm{s}^L \int \mathrm{d}\Pi_L f^{(L)},
\end{equation}
where $f^{(L)}$ is integrable for any $L$? If that's the case, since $f^{(L)}$ is integrable, the integral can be performed numerically using Monte Carlo techniques. The traditional approach involves the analytic computation of loop integrals (see ref.~\cite{2021} and references therein for a review of the state-of-the-art) using dimensional regularisation and the numerical evaluation of appropriately subtracted phase-space integrals~\cite{Currie:2016bfm,Czakon:2010td,Boughezal:2015dra,Somogyi:2005xz,DelDuca:2016ily,Grazzini:2017mhc,Cieri:2018oms,Boughezal:2016wmq,Cacciari:2015jma,Caola:2017dug,Herzog:2018ily,Magnea:2018hab}. Yet another approach is to also compute loop integrals numerically after performing subtraction of their singularities~\cite{Anastasiou_2019,Anastasiou_2021,Capatti_2020,Ma_2020,Runkel_2019,Runkel_2020,Hern_ndez_Pinto_2016,Becker_2010,Becker_2012,Gong_2009}. In both cases, loop and phase-space integrals are treated as independent and separately diverging pieces of the computation, and the singularities are expected to disappear after all the independent pieces are combined together. Here we will present a different method, that instead aims to directly compute interference diagrams and regulate their infrared singularities \textit{without the need of any subtraction technique nor dimensional regularisation}.

We start by analysing precisely the integral representation of interference diagrams, e.g.

\begin{minipage}{13.52cm}
\begin{figure}[H]
\centering
\includegraphics[width=13.5cm]{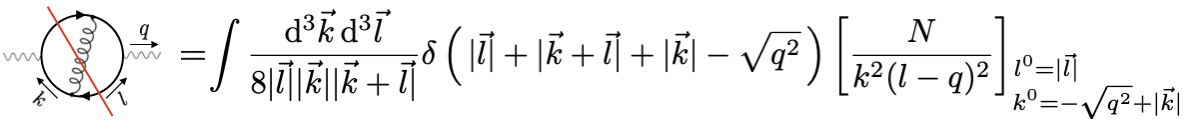}
\end{figure}
\end{minipage}%
\begin{minipage}{1cm}
\begin{center}
\vspace{0.2cm}
  (7)  
\end{center}
\end{minipage}
\vspace{0.0cm}

\begin{minipage}{13.52cm}
\begin{figure}[H]
\centering
\hspace{-2.15cm}\includegraphics[width=11.7cm]{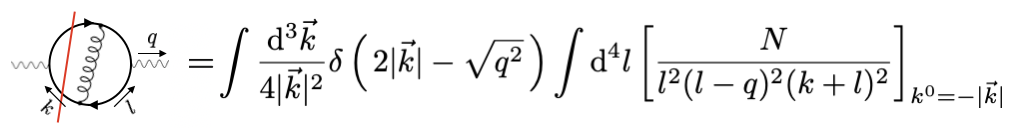}
\end{figure}
\end{minipage}%
\begin{minipage}{1cm}
\begin{center}
\vspace{0.2cm}
  (8)  
\end{center}
\end{minipage}
\vspace{0.25cm}

\noindent Recall that $q$ is set to take its rest frame value and thus $\vec{q}=0$. The integration measure of the two interference diagrams eq.~$(7)$ and eq.~$(8)$ differs in two aspects: a) the real emission, eq.~$(7)$, has an integration in $\mathrm{d}^3\vec{l}$, while the virtual contribution, eq.~$(8)$, is integrated over $\mathrm{d}^4l$, b) the argument of the remaining Dirac delta function enforcing the sum of the on-shell energies of cut particles to be the same as the invariant mass of the decaying particle. Then, our objective can be schematically represented as follows: starting from $I_{\mathrm{dt}}$ as defined in eq.~$(5)$ as a sum of four, distinct, integrals, with different integrands and different integration measures, we would like to represent it as
\begin{equation}\tag{9}
    I_{\mathrm{dt}}=\sum_{i=1}^4 \int \mathrm{d}\Pi_i f_i=\int \mathrm{d}\Pi \sum_{i=1}^4 f_i.
\end{equation}
This is what we mean by "aligning the integration measure": the integral on the right hand side of eq.~$(9)$ factorises an integration measure that is common to all interference diagrams, and the integrand is the sum of four functions, each corresponding to a distinct interference diagram.

At its core, the Local Unitarity~\cite{Capatti_2021,Soper_1998,Soper_2000} procedure consists of three steps:
\begin{enumerate}
    \item[$\mathrm{i)}$] Assign the same loop momentum routing to all the interference diagrams with the same underlying forward scattering diagram.
    
    \item[$\mathrm{ii)}$] Use LTD (or cLTD, or TOPT) in order to perform all the loop energy integrations.
    
    \item[$\mathrm{iii)}$] Use the \textit{causal flow} to solve all the energy conserving Dirac delta functions using a newly introduced variable that is common to all interference diagrams.

\end{enumerate}

\subsection{Integrating energies}

The first obstacle to aligning the integration measure of different interference diagrams lies in the different dimensionality of loop and phase-space integrals. This issue is specific to covariant perturbation theory, in which amplitudes are expressed as integrals over Minkowski space. 


One way to solve the issue is to work within the Loop-Tree Duality formalism~\cite{Aguilera_Verdugo_2020,Aguilera_Verdugo_2021,Bierenbaum_2010,bobadilla2021lotty,Capatti_2019,capatti2020manifestly,Catani_2008,Runkel_2019,Runkel_2020,Sborlini_2021} (or, alternatively, in Time-Ordered Perturbation Theory). Broadly speaking, loop energy integrations can be performed analytically using the residue theorem: 
\begin{equation}\tag{10}
    \int \prod_{i=1}^L \frac{\mathrm{d}^4 k_i}{(2\pi)^4} \frac{N}{\prod_{j\in\mathbf{e}}(q_j^2-m_j^2-\mathrm{i}\epsilon)}=\int \prod_{i=1}^L \frac{\mathrm{d}^3 \vec{k}_i}{(2\pi)^3} f_{\mathrm{ltd}}.
\end{equation}
 A closed formula for $f_{\mathrm{ltd}}$, for generic numerators and topologies, is provided in ref.~\cite{Capatti_2019}. $f_{\mathrm{ltd}}$ can itself be represented diagrammatically, by using the same definition of Cutkosky cuts presented for interference diagrams. For the example at hand, we apply LTD to the $l$-loop triangle of the virtual contribution in eq.~(7):
 
\begin{minipage}{13.52cm}
\begin{figure}[H]
\centering
\includegraphics[width=13.5cm]{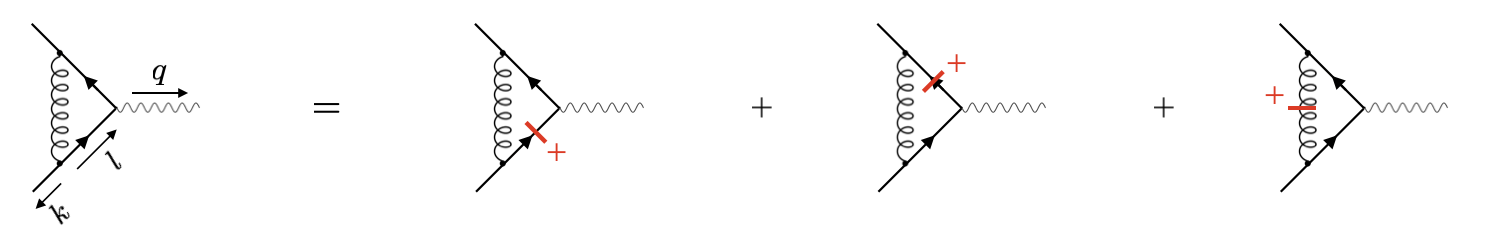}
\end{figure}
\end{minipage}%
\begin{minipage}{1cm}
\begin{center}
\vspace{0.2cm}
  (11)  
\end{center}
\end{minipage}
\vspace{0.0cm}

\noindent Translating this result to the interference diagram, we obtain

\begin{minipage}{13.52cm}
\begin{figure}[H]
\centering
\hspace{-0.05cm}\includegraphics[width=13.5cm]{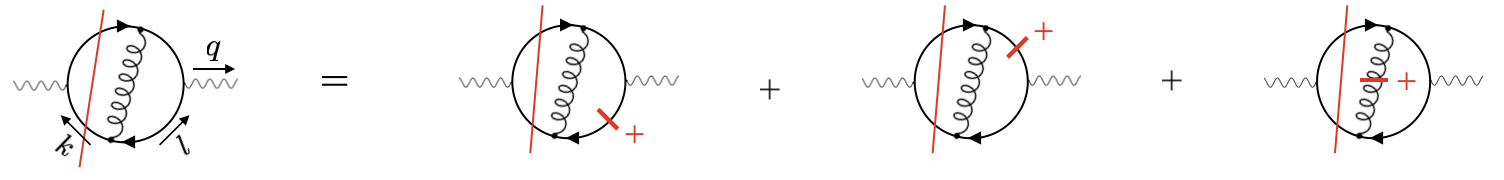}
\end{figure}
\end{minipage}%
\begin{minipage}{1cm}
\begin{center}
\vspace{0.2cm}
  (12)  
\end{center}
\end{minipage}
\vspace{0.25cm}

\noindent After the application of LTD, the number of cut particles on the rhs of eq.~$(12)$ is the same as the number of particles cut in the real emission eq.~$(7)$. That is: the number of particles that are constrained to live on the mass-shell is the same for all interference diagrams, after the application of LTD. In other words, after the application of LTD the integration space has the same dimensionality for virtual and real contributions, which loosens the distinction between phase-space integrals and loop integrals.

One fundamental property of the cut diagrams is that, after the deletion of the cut edges, one is left with a connected tree, also known as spanning tree. This is where the name Loop-Tree Duality originates. On a conceptual level, this shows that loop integrations amount to summing over all the tree processes that can be embedded within the virtual loops (of course, contrary to particles crossed by Cutkosky cuts, particles crossed by an LTD cut can be infinitely energetic).

Furthermore, the diagrammatic description also pairs each cut with a sign, which in this case appears to be $+$ for all the cuts. These signs prescribes which one of the two on-shell energy solutions is to be taken. This sign is called the \textit{energy flow} associated with the cut.

The extension of LTD from one-loop to multi-loop topologies is non-trivial. In order to shed light on the subleties that arise from the multi-loop extension of LTD, we report the LTD representation of the double triangle

\begin{minipage}{13.52cm}
\begin{figure}[H]
\centering
\hspace{-0.8cm}\includegraphics[width=14cm]{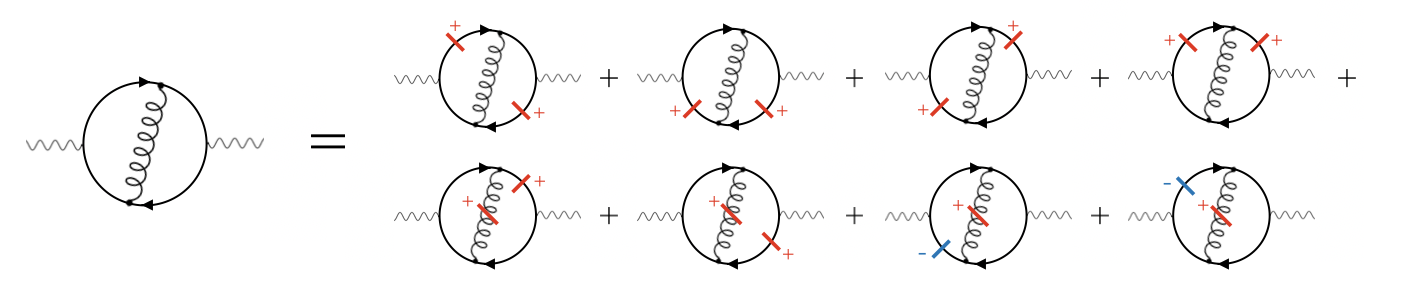}
\end{figure}
\end{minipage}%
\begin{minipage}{1cm}
\begin{center}
\vspace{0.2cm}
  (13)  
\end{center}
\end{minipage}
\vspace{0.25cm}

While these cut diagrams still exhibit the property of being spanning trees after the deletion of the edges that are being cut, the energy flow appears less trivial to derive. The rigorous derivation of the energy flow is at the origin of the complications arising from the extension of the Loop-Tree Duality to multi-loop topologies. It is derived by carefully disentangling  the interplay between momentum conservation and causality, i.e. the $\mathrm{i}\epsilon$ prescription~\cite{Capatti_2019}.

\subsection{The causal flow}
\label{sect:causal_flow}

The second obstacle in alligning the integration measure of distinct interference diagrams lies in the Dirac delta function enforcing the sum of the on-shell energies of the particles crossed by the Cutkosky cut to be the same as the energy of the deying particle. The argument of such Dirac delta functions differs across interference diagrams. After the application of LTD, row $(b)$ of eq.~$(3)$ is rewritten as
\begin{equation}\tag{14}
    I_{\mathrm{dt}}=\int \mathrm{d}^3 \vec{k}\, \mathrm{d}^3 \vec{l} \, \, \left[\delta\left( \, |\vec{k}|+|\vec{l}|+|\vec{k}+\vec{l}|-\sqrt{q^2} \, \right)  f^{\text{r}}\left(\vec{k}, \,\vec{l} \, \right)+\delta\left( \, 2|\vec{k}|-\sqrt{q^2} \, \right)  f^{\text{v}}\left(\vec{k}, \,\vec{l} \, \right)+(\,\vec{k}\rightarrow \vec{l}\, \,)\right]
\end{equation}
\noindent where $f^\mathrm{r}$ and $f^\mathrm{v}$ are functions that are infinitely differentiable almost everywhere, and can be obtained from the comparison with eq.~(6) and eq.~(7). The integration measure now differs only in the Dirac delta functions enforcing conservation of on-shell energies. The surface described by the zeros of the argument of one of such Dirac delta functions constitutes a connected physical threshold. Explicitly, the four thresholds are in one to one correspondence with the interference diagrams summing up to $I_{\mathrm{dt}}$, are:
\begin{align}\tag{15}
\eta_1=2|\vec{k}|-\sqrt{q^2}=0, \quad  \eta_2=2|\vec{l}|-\sqrt{q^2}=0, \quad \eta_3=\eta_4=|\vec{k}|+|\vec{k}+\vec{l}|+|\vec{l}|-\sqrt{q^2}=0.
\end{align}
The first two equations correspond to the two virtual contributions, whereas the last two corresponds to both real emissions (in the rest frame). They are bounded and convex surfaces. We now solve the Dirac delta functions without breaking the alignment of the integration measure. This is a difficult step, and the origin of the complicated mappings that are used in the traditional subtraction of phase-space integrals, which usually require some form of sectoring. Solving any two deltas using two different variables would automatically break the alignment of the measure, and spoil the benefits gained from having factorised the $\mathrm{d}^3 \vec{k}\, \mathrm{d}^3 \vec{l}$ integration measure. 

In ref.~\cite{Soper_1998}, D.Soper proposed a very effective trick that allows to elegantly solve this problem. The generalisation of this trick to arbitrary order and topologies is called the \textit{causal flow}~ \cite{Capatti_2021}. The basic idea is that one should introduce a new variable that can be used to solve all the deltas simultaneously, thus mantaining the alignment of the measure. Here we present it in its simplest form. Let us focus on one of the interference diagrams, and introduce a resolution of the identity $1=\int \mathrm{d}t \, h(t) $,
\begin{equation}\tag{16}
I^{\mathrm{v}}=\int \mathrm{d}^3 \vec{k} \, \mathrm{d}^3 \vec{l} \, \, \delta\left(2|\vec{k}|-\sqrt{q^2}\right)  f^{\mathrm{v}}\left(\vec{k}, \, \vec{l} \,\right)=\int \mathrm{d}^3 \vec{k} \, \mathrm{d}^3 \vec{l} \int \mathrm{d}t \, h(t) \, \, \delta\left(2|\vec{k}|-\sqrt{q^2}\right)  f^{\mathrm{v}}\left(\vec{k}, \, \vec{l} \,\right).
\end{equation}
We then consider the following rescaling change of variables $(\vec{k},\vec{l})\rightarrow\phi(t)= e^t(\vec{k},\vec{l})$. $\phi(t)$ is a one-parameter group. Under this change of variables, the spatial momentum carried by any particle is rescaled by $e^t$. As a result, $e^t$ will now appear in the argument of the delta
\begin{equation}\tag{17}\label{scaling_integral}
    I^{\mathrm{v}}=\int \mathrm{d}^3 \vec{k} \, \mathrm{d}^3 \vec{l} \, \mathrm{d}t \, e^{6t}h(t) \, \, \delta\left(2e^t|\vec{k}|-\sqrt{q^2}\right)  f^{\mathrm{v}}\left(e^t\vec{k}, \, e^t\vec{l} \,\right).
\end{equation}
This allows to solve the Dirac delta using the variable $t$. The value of $t$ is then set to the unique solution
\begin{equation}\tag{18}
    t_\mathrm{v}^\star=\log\left(\frac{\sqrt{q^2}}{2|\vec{k}|}\right),
\end{equation}
so that the resulting integral will feature no delta anymore:
\begin{equation}\tag{19}\label{scaling_integral}
    I^{\mathrm{v}}=\int \mathrm{d}^3 \vec{k} \, \mathrm{d}^3 \vec{l}\, \left[\frac{e^{5t^\star_\mathrm{v}}h(t^\star_\mathrm{v})}{2|\vec{k}|}\right]  f^{\mathrm{v}}\left(e^{t^\star_\mathrm{v}}\vec{k}, \, e^{t^\star_\mathrm{v}}\vec{l} \,\right).
\end{equation}
Eq.~(19) also factors in the Jacobian coming from the delta, $2|\vec{k}|e^{t^\star_\mathrm{v}}$. It is important that we have not used the spatial integrations in $\vec{k}$ or $\vec{l}$ to solve the delta. Using the variable $t$ allows to democratically solve all the deltas for any interference diagrams in the \textit{same} variable. We started with an integral in eq.~(11) whose measure is $\mathrm{d}^3 \vec{k}\, \mathrm{d}^3 \vec{l}$ and we solved the delta obtaining an integral, in eq.~(14), which factorises the same exact integration measure.

The rescaling change of variable $\phi(t)=e^t(\vec{k},\vec{l})$ satisfies the flow equation $\partial_t \phi(t)=\kappa(\phi(t))$, with $\kappa(\phi(t))=\phi(t)$ and with boundary condition $\phi(0)=(\vec{k},\vec{l})$. The vector field $\kappa$, which in this case is very simple, satisfies a very important property.  Observe that $\kappa_k\cdot\nabla_k\eta_i+\kappa_l\cdot\nabla_l\eta_i>0$ everywhere and thus, more specifically, whenever $\eta_i(\vec{k})=0$. That is: the vector field $\kappa$ has positive projection on the normal of the threshold surfaces $\eta_i$ at any point located on the surface itself, hence the name \textit{causal flow}. Such a property is fundamental to construct the correct change of variable. The proper generalisation of the change of variable for general $\mathrm{N}^k\mathrm{LO}$, $N$ to $M$ processes~\cite{Capatti_2021} is indeed the solution of a flow equation, whose defining vector field $\kappa$ has positive projection on the normal vector of all the threshold surfaces of the forward-scattering diagram~\cite{Capatti_2020}.

\subsection{Local Unitarity representation}

The procedure used in sect.~\ref{sect:causal_flow} can be applied to any of the other contributions building up $I_{\mathrm{dt}}$, yielding an integral whose integration measure is simply $\mathrm{d}^3\vec{k}\mathrm{d}^3\vec{l}$, with no extra Dirac delta functions to be solved:

\begin{equation}\tag{20}\label{lu_rep}
    I_{\mathrm{dt}}=\int\mathrm{d}^3 \vec{k} \mathrm{d}^3 \vec{l} \, \, \left[  \sum_{i=1}^4 \frac{h(t_i^\star) e^{5 t_i^\star}}{\eta_i(\vec{k}, \,\vec{l} \,)+\sqrt{q^2}} f^i\left(e^{t^\star_i}\vec{k}, \, e^{t^\star_i}\vec{l} \,\right)\right],
\end{equation}
where we identify 
\begin{equation}\tag{21}
f^1(\vec{k}, \, \vec{l} \,)=f^\mathrm{v}(\vec{k}, \, \vec{l} \,), \ \ f^2(\vec{k}, \, \vec{l} \,)=f^\mathrm{v}(\vec{l}, \, \vec{k} \,), \ \ f^3(\vec{k}, \, \vec{l} \,)=f^\mathrm{r}(\vec{k}, \, \vec{l} \,), \ \ f^4(\vec{k}, \, \vec{l} \,)=f^\mathrm{r}(\vec{l}, \, \vec{k} \,),
\end{equation}
and
\begin{equation}\tag{22}
    t^\star_i=\log\left(\frac{\sqrt{q^2}}{\eta_i(\vec{k}, \,\vec{l} \,)+\sqrt{q^2}}\right).
\end{equation}
The integration measure of all interference diagrams is thus fully aligned in a \textit{causal} way. The next task is to testablish if the resulting integral is well-defined, and if the integrand itself is Lebesgue integrable. This is the case, although it is non-trivial to show it, especially for more complicated processes: we provided a general proof of the absence of infrared singularities for generic processes in~\cite{Capatti_2021}. In summary:

\vspace{0.5cm}
\textbf{The integrand resulting from the alignment of the measure operated through LTD and the causal flow is locally free of non-integrable infrared singularities.}
\vspace{0.5cm}

Since the method is fully local, the generalisation of the alignment procedure to \textit{generic observables} is trivial, provided they satisfy the \textit{IR-safety condition}. In practice, this is achieved by multiplying each summand of eq.~\eqref{lu_rep} by its corresponding observable (with rescaled inputs) to obtain a quantity that is still locally free of IR-singularities. In other words, one can operate the substitution $f^i\rightarrow f^i \mathcal{O}^i$ where $\mathcal{O}^i$ is any IR-safe function of the external momenta of the $i$-th cut. $\mathcal{O}^i$ is then function of the rescaled final-state momenta.

\subsection{LU as local residues of the supergraph}

The procedure of solving the deltas by using the causal flow variable $t$ suggests that the zeros of $\eta_i$ can be seen as simple poles in the variable $t$ along the causal flow. We will use this to construct a generic and powerful expression for the LU representation of a cross-section. Let us consider again the example put forward in the previous section. We start with the LTD representation of the double triangle eq.~(11), and rescale the loop momenta $\vec{k},\vec{l}$:

\begin{minipage}{13.52cm}
\begin{figure}[H]
\centering
\includegraphics[width=9cm]{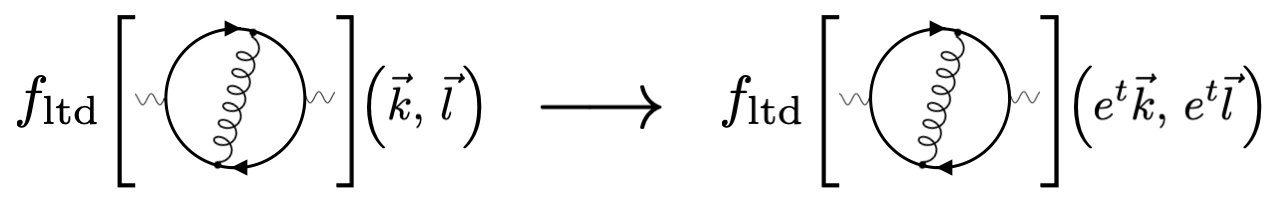}
\end{figure}
\end{minipage}%
\begin{minipage}{1cm}
\begin{center}
\vspace{0.2cm}
  (23)  
\end{center}
\end{minipage}
\vspace{0.25cm}

 We can now study the singular behaviour of this object. The singular thresholds of the LTD representation of the double-triangle forward-scattering diagram are in one-to-one correspondence with Cutkosky cuts~\cite{Cutkosky:1960sp}, and thus with the $\eta_i$, as defined in eq.~$(15)$. Thus, aside from any phase-space location corresponding to a particle becoming soft, the points at which the LTD representation of the double-triangle forward-scattering diagram is singular are entirely described implicitly by the equations $\eta_i=0$, $i=1,...,4$. The behaviour of the right hand side of eq.~$(21)$ close to such thresholds (and at any other point in space) can be studied in the parameter $t$. For any fixed $\vec{k}$ and $\vec{l}$, $\eta_i(e^t\vec{k},e^t\vec{l})$ admits one unique zero in the variable $t$, and the expansion around such zero reads
\begin{equation}\tag{24}
    \eta_i(e^t\vec{k},e^t\vec{l})=(t-t^\star_i)\left[\eta_i(\vec{k}, \,\vec{l} \,)+\sqrt{q^2}\right]e^{t^\star_i}+\mathcal{O}((t-t^\star_i)^2), \quad i=1,...,4,
\end{equation}
that is, the integrand only has simple poles in the variable $t$.
Notice that the first non-zero coefficient in the expansion matches the Jacobian factor induced by the Dirac delta function introduced in eq.~(17). Finally, let us consider a weighted sum of the residues in $t$ of eq.~(18):

\begin{minipage}{13.52cm}
\begin{figure}[H]
\centering
\includegraphics[width=9cm]{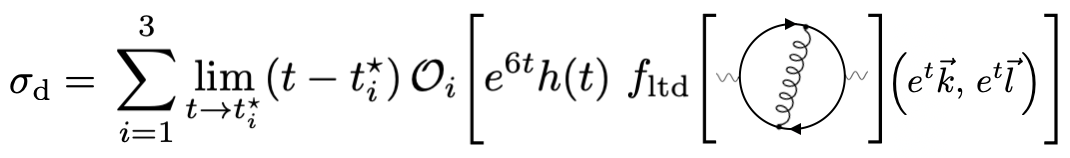}
\end{figure}
\end{minipage}%
\begin{minipage}{1cm}
\begin{center}
\vspace{0.2cm}
  (25)  
\end{center}
\end{minipage}
\vspace{0.25cm}

\noindent Observe that the sum runs over three limits in $t$ instead of four, due to $\eta_3$ being the same as $\eta_4$ in the rest frame. An in-depth comparison with eq.~(17) reveals that 
\begin{equation}\tag{26}
    I_{\mathrm{dt}}=\int \mathrm{d}^3 \vec{k} \mathrm{d}^3 \vec{l} \, \sigma_{\mathrm{d}}.
\end{equation}
The consequences of this equality cannot be understated: a locally IR-finite representation of differential cross-sections is obtained by summing over the local residues in the variable $t$ along the flow lines of $\phi(t)$. This reproduces Cutkosky's well-known result~\cite{Cutkosky:1960sp}, that interference diagrams can be obtained from the thresholds of forward-scattering diagrams. This correspondence, in Local Unitarity, is established at the local level, which results in the local cancellation of IR-singularities.

\section{Conclusion}

We have presented the main aspects of the Local Unitarity representation, including a simplified description of the mathematical tools that its derivation requires. Such tools can be shown to generically apply at any perturbative order and for physical processes with arbitrary number of scales. If one also includes a systematic and local regularisation of UV singularities as well as of physical, integrable thresholds~\cite{Capatti_2020}, a clear path towards automation emerges.

\section*{Acknowledgements}
I would like to thank Valentin Hirschi and Ben Ruijl for their comments on this proceeding.

\bibliography{biblio.bib}




\nolinenumbers

\end{document}